\documentclass[%
 reprint,
superscriptaddress,
 amsmath,amssymb,
 aps,
pra,
]{revtex4-2}

\usepackage{amsmath}
\usepackage{amssymb}
\usepackage{amsthm}
\usepackage{physics}
\usepackage{graphicx}
\usepackage{dcolumn}
\usepackage{bm}
\usepackage{hyperref}

\theoremstyle{definition}

\usepackage{xcolor}

\begin{document}

\preprint{APS/123-QED}

\title{Long-distance device-independent quantum key distribution with standard optics tools}

\author{Makoto Ishihara}
 \email{llmakomako.arg1076@keio.jp}
 \affiliation{%
 Department of Electronics and Electrical Engineering, Keio University, 3-14-1 Hiyoshi, Kohoku-ku, Yokohama 223-8522, Japan
}%
\author{Anthony Brendan}
\email{anthony.brendan@keio.jp}
\affiliation{%
 Department of Electronics and Electrical Engineering, Keio University, 3-14-1 Hiyoshi, Kohoku-ku, Yokohama 223-8522, Japan
}%
\author{Wojciech Roga}%
 \email{wojciech.roga@keio.jp}
 \affiliation{%
 Department of Electronics and Electrical Engineering, Keio University, 3-14-1 Hiyoshi, Kohoku-ku, Yokohama 223-8522, Japan
}%
\author{Ulrik L. Andersen}%
 \email{ulrik.andersen@fysik.dtu.dk}
 \affiliation{%
 Center for Macroscopic Quantum States (bigQ), Department of Physics, Technical University of Denmark, 2800 Kongens Lyngby, Denmark
}%
\author{Masahiro Takeoka}%
 \email{takeoka@elec.keio.ac.jp}
\affiliation{%
 Department of Electronics and Electrical Engineering, Keio University, 3-14-1 Hiyoshi, Kohoku-ku, Yokohama 223-8522, Japan
}%
\affiliation{%
 National Institute of Information and Communications Technology (NICT), Koganei, Tokyo 184-8795, Japan
}%

\date{\today}

\begin{abstract}
Device-independent quantum key distribution (DI-QKD) enables information-theoretically secure key exchange between remote parties without any assumptions on the internal workings of the devices used for its implementation. However, its practical deployment remains severely constrained by the need for loophole-free Bell inequality violations, which are highly susceptible to losses and detection efficiencies. In this paper, we propose two long-distance DI-QKD protocols based on a heralding scheme using single-photon interference. Our protocols consist of only standard quantum optics tools such as two-mode squeezed states, displacement operations and on-off detectors, making them experimentally accessible. To further enhance robustness against realistic imperfections, we integrate a classical noisy preprocessing technique during post-processing. We calculate key rates of the protocols by numerical optimization and show the supremacy of this implementation over existing protocols in terms of communication distances.

\end{abstract}

\maketitle


\section{\label{sec:introduction}INTRODUCTION}
Quantum key distribution (QKD) realizes information-theoretically secure key exchange among remote parties~\cite{Bennett2014, Ekert1991,Xu2020, Pirandola2020}. The security of conventional QKD, however, requires the assumption that all the devices used during the implementation must be completely characterized and fully trustworthy, which is a condition that is difficult to meet in realistic implementation. Device-independent QKD (DI-QKD) resolves this problem by leveraging violations of Bell inequalities~\cite{Acin2007, Pironio2009} to certify security independently of the internal workings of the devices. By observing a loophole-free Bell violation, DI-QKD ensures that only quantum systems capable of such correlations could have produced the observed statistics. While most research on DI-QKD remains theoretical~\cite{Zapatero2023, Primaatmaja2023}, several proof-of-principle experimental demonstrations have recently been achieved~\cite{Nadlinger2022, Zhang2022, Liu2022}.

Despite its strong security guarantees, DI-QKD is notoriously sensitive to experimental imperfections such as channel loss and detector inefficiencies. In particular, achieving a loophole-free Bell violation requires high overall efficiency, which significantly limits the communication distance. For instance, the record distance for a fully photonic DI-QKD experiment is 220 meters~\cite{Liu2022}, which falls short of practical requirements.

To overcome this challenge, DI-QKD protocols employing heralding schemes have been proposed~\cite{Gisin2010, Seshadreesan2016, Zapatero2019, Tsujimoto2020, Kolodynski2020, Mycroft2023, Steffinlongo2024, Alwahaibi2025}. Such heralding schemes mitigate the effect of channel loss by treating it as a probabilistic preparation of entanglement, effectively extending the communication distance significantly. Among these approaches, heralding via single-photon interference, as employed in twin-field QKD~\cite{Lucamarini2018}, is promising. It acts as a single-hop quantum repeater and enables significantly longer communication distances compared to protocols based on two-photon interference~\cite{Lo2012}. In the context of DI-QKD and the loophole-free Bell test, single-photon interference based protocols are proposed with two-mode squeezed vacua~\cite{Mycroft2023}, single-photon entangled states~\cite{Steffinlongo2024}, and their hybrids~\cite{Alwahaibi2025}.

In this work, we propose two practical long-distance DI-QKD protocols that utilize heralding based on single-photon interference. 
Our protocols rely exclusively on standard quantum optical components such as two-mode squeezed states generated from spontaneous parametric down conversion, displacement operations, and photon detectors discriminating only zero or non-zero photons (on-off detectors).  
This is contrasted to other recent protocols~\cite{Mycroft2023, Steffinlongo2024, Alwahaibi2025} which require more demanding elements such as single-photon emitters and the photon-number-resolving detectors.
We calculate key rates of our protocols using the numerical optimization approach developed in Refs.~\cite{Navascues2008, Masanes2011, NietoSilleras2014, Bancal2014, Brown2024} and enhance the protocols' tolerance to detection imperfections via noisy preprocessing~\cite{Ho2020}, where legitimate users deliberately introduce controlled noise into their key bits to reduce an eavesdropper's information. Our analysis shows that the proposed protocols enable secure key distribution over longer distances than existing DI-QKD protocols, using only experimentally accessible resources.
We also discuss the minimum required detector efficiency for our protocols and the effect of detector dark counts.


The paper is structured as follows. In Sec. \ref{section:protocol}, we introduce the proposed DI-QKD protocols. In Sec.~\ref{sec:ProbabilityDistribution}, we derive the relevant joint probability distribution between remote parties which are used for key rate calculation. Then, we explain how to calculate key rates of our protocols in Sec.~\ref{sec:Keyrate}. We provide results of the key rate calculation in Sec.~\ref{sec:results}, and we conclude in Sec.~\ref{sec:conclusion}.

\section{PROTOCOLS}\label{section:protocol}
\begin{figure*}[htbp]
 \centering
 \includegraphics[keepaspectratio, scale=0.45]{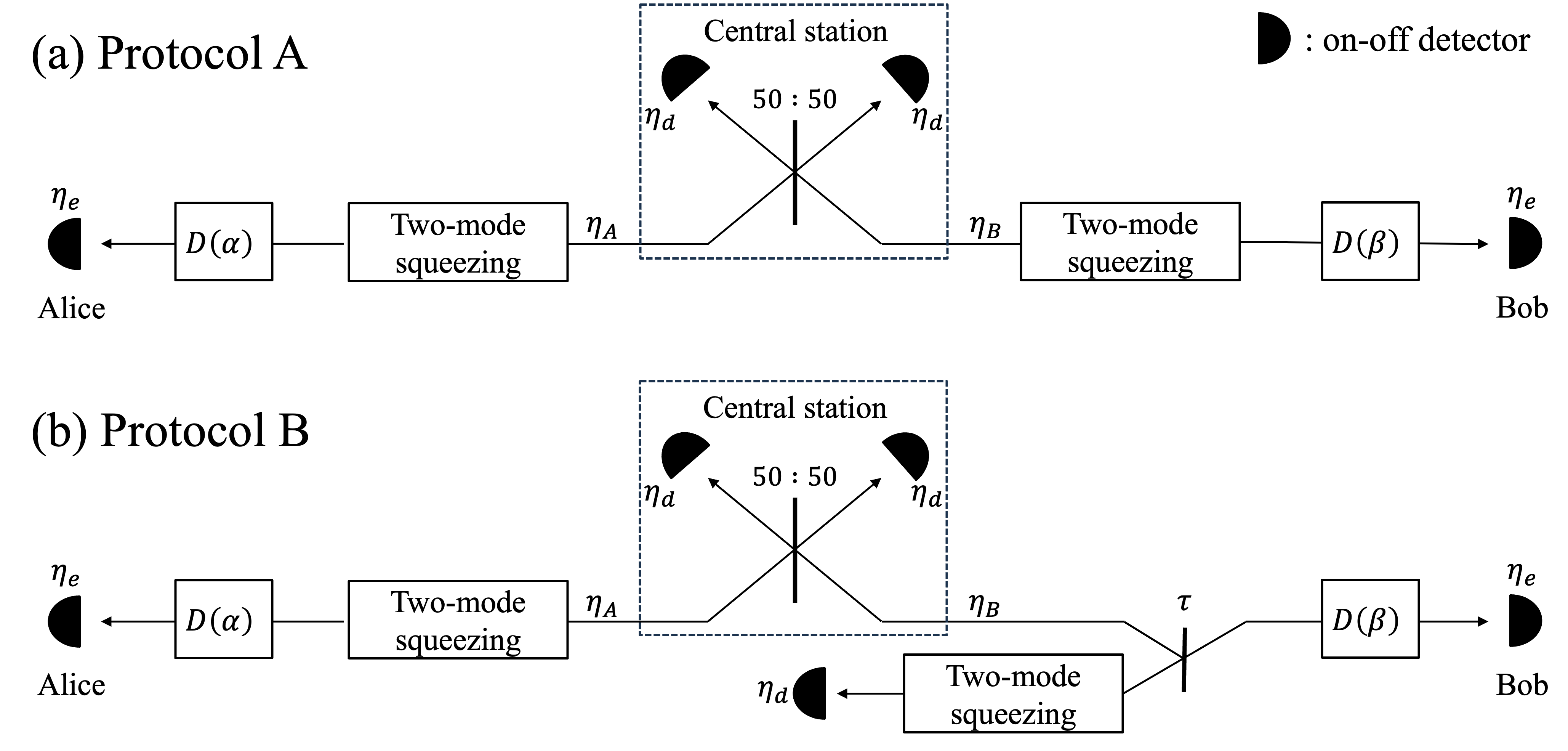}
 \caption{Schematics of our DI-QKD protocols. (a) Protocol A. Alice and Bob prepare two-mode squeezed states and transmit one-mode to the central station through pure-loss channels. At the central station, these modes are interfered with a 50:50 beamsplitter and detected with on-off detectors. Alice and Bob perform displacement operations on their remaining modes followed by measurements using on-off detectors. Alice and Bob share a secret key by using events where exactly one of the two on-off detectors clicks. (b) Protocol B. Bob prepares a single-photon entanglement by using a two-mode squeezed state, a heralding detector and a beamsplitter, and sends one-mode of the entanglement to the central station. Alice and Bob share a secret key from events where only one of the two on-off detectors at the central station and the heralding detector click simultaneously.}
 \label{fig:LDDIQKDsetup}
\end{figure*}

We present two DI-QKD protocols, referred to as Protocol A and Protocol B, illustrated in Fig.~\ref{fig:LDDIQKDsetup}. In Protocol A (Fig.~\ref{fig:LDDIQKDsetup} (a)), two legitimate parties, Alice and Bob, aim to share a secure key. Alice (Bob) prepares a two-mode squeezed state with mean photon number $\bar{n}_A$ ($\bar{n}_B$) and each transmits one-mode to a central station through a pure-loss channel with transmissivity $\eta_A$ ($\eta_B$). The central station is composed of a 50:50 beamsplitter and two on-off detectors with detection efficiencies $\eta_d$ and dark count probabilities $p_d$. A successful heralding event occurs when exactly one of the two detectors registers a click. Upon successful heralding event, Alice and Bob perform displacement operations on their remaining modes followed by measurements using on-off detectors with detection efficiencies $\eta_e$ and dark count probabilities $p_d^e$. They postselect successful events and use these events for key generation or Bell test rounds.

Alice (Bob) selects between two (three) different displacement settings denoted by $x \in \{ 0, 1 \}$ ($y \in \{ 0, 1, 2 \}$). Let $a \in \{ 0, 1\}$ ($b \in \{ 0, 1 \}$) denote Alice's (Bob's) measurement outcomes where we assign ``0'' to no click at the on-off detector and ``1'' to a click. The protocol classifies the postselected successful events into two categories. Key generation rounds correspond to events where Alice selects $x=0$ and Bob selects $y=2$, and the Bell test rounds correspond to all other combinations of $x \in \{ 0, 1\}$ and $y \in \{0, 1, 2\}$. Let $P(a,b|x,y)$ denote the joint probability distribution where Alice (Bob) chooses $x$ ($y$) and obtains $a $ ($b$).

To further enhance security, we employ noisy preprocessing during key generation. Specifically, Alice randomly flips her bits with a probability $p_n$. Although this reduces the correlation between Alice and Bob, it also diminishes information of an eavesdropper, Eve, potentially resulting in a net gain in security.

In Protocol B (Fig.~\ref{fig:LDDIQKDsetup} (b)), the state preparation on Bob's side is modified. Instead of using a two-mode squeezed state, Bob generates a single-photon entangled state, created via a two-mode squeezed state with mean photon number $\bar{n}_B$, a heralding on-off detector with efficiency $\eta_d$ and dark count probability $p_d$, and a beamsplitter with transmissivity $\tau$. When the heralding detector and one of the two central station detectors click simultaneously, the event is marked as successful. The remaining steps of the protocol are equivalent to those in Protocol A. Note that this configuration was first proposed for long-distance QKD~\cite{Winnel2021} and later for the loophole-free Bell test \cite{Alwahaibi2025}.

\section{DERIVING JOINT PROBABILITY DISTRIBUTION}\label{sec:ProbabilityDistribution}
In this section, we derive the probability distribution $P(a, b|x, y)$ for measurement outcomes $a$ and $b$ given the displacement settings $x$ and $y$. First, we focus on Protocol A. Let us denote a system of $N$ bosonic modes by their annihilation and creation operators $\{\hat{a}_i, \hat{a}^\dagger_i  \}_{i=1}^N$ which satisfy the commutation relations $[\hat{a}_i, \hat{a}^\dagger_j] = \delta_{ij}$. By using these operators, we define the quadrature field operators $\{\hat{q}_i, \hat{p}_i \}_{i=1}^N$ which satisfy $\hat{q}_i = \hat{a}_i + \hat{a}^\dagger_i$ and $\hat{p}_i = i(\hat{a}^\dagger_i - \hat{a}_i)$.
In the $N$-mode bosonic system, a quantum state $\rho$ is characterized by its characteristic function
\begin{equation}
    \chi(x) = \text{Tr} [ \rho \mathcal{W} (x)],
\end{equation}
where $\mathcal{W}$ is a Weyl operator defined as
\begin{equation}
    \mathcal{W} (x) = \exp [-i x^T r],
\end{equation}
where $x \in \mathbb{R}^{2N}$ is a real vector and $r = [\hat{q}_1, \hat{p}_1, \ldots, \hat{q}_N, \hat{p}_N]^T$.

The covariance matrix of a two-mode squeezed state with mean photon number $\bar{n}$ is as follows~\cite{Weedbrook2012, Takeoka2015}
\begin{equation}
    \Gamma^{\text{TMSV}} (\bar{n}) = \left[ \begin{array}{cc}
    v I_2 & \sqrt{v^2-1} Z\\
    \sqrt{v^2-1} Z & v I_2
    \end{array}
    \right],
\end{equation}
where $I_n$ is the $n \times n$ identity matrix and 
\begin{equation}
    Z = \left[  
    \begin{array}{cc}
        1 & 0 \\
        0 & -1
    \end{array}
    \right].
\end{equation}
The parameter $v$ corresponds to squeezing level of the two-mode squeezed state, and it relates to the mean photon number of the state $\bar{n}$ as $v = 1 + 2\bar{n}$. Here, we omit an argument for a displacement vector since it is always a zero vector. The total initial covariance matrix for the system, including both signal and environmental modes, is then:
\begin{equation}
    \Gamma^{\text{initial}} = \Gamma_{AA'}^{\text{TMSV}} (\bar{n}_A) \oplus \Gamma_{BB'}^{\text{TMSV}} (\bar{n}_B) \oplus I_{E_A^1E_A^2F_AE_B^1E_B^2F_B},
\end{equation}
where modes $A$ and $B$ are detected by Alice and Bob, modes $A'$ and $B'$ are sent to the central station, $E_A^1$ and $E_B^1$ are environmental modes for the pure-loss channels, $E_A^2$ and $E_B^2$ are environmental modes for inefficiencies of the on-off detectors at the central station, and $F_A$ and $F_B$ are environmental modes for the on-off detectors of Alice and Bob, respectively. We model each pure-loss channel with transmissivity $\eta$ as a beamsplitter with vacuum in one of the input modes. Similarly, the inefficiencies in on-off detectors can be treated as effective losses prior to detection. Importantly, losses in Alice's and Bob's on-off detectors can be modeled before the displacement operations. Since detector losses correspond to a linear transformation, they can be absorbed into the effective displacement amplitudes without changing the fundamental physics. The operation of the beamsplitter with transmissivity $\eta$ on the quadratures can be expressed as follows
\begin{equation}
    S^{\text{BS}} (\eta) = \left[
    \begin{array}{cc}
        \sqrt{\eta}\, I_2 & \sqrt{1-\eta}\, I_2 \\
        -\sqrt{1-\eta}\, I_2 & \sqrt{\eta}\, I_2
    \end{array}
    \right].
\end{equation}
Then the operation of the pure-loss channels is
\begin{equation}
    S^{\text{channel}} = S^{\text{BS}}_{A'E_A^1} (\eta_A) \oplus S^{\text{BS}}_{B'E_B^1} (\eta_B) \oplus I_{AE_A^2F_ABE_B^2F_B}
\end{equation}
We obtain a state after the channel transmission by applying this operation to the initial state
\begin{equation}
    \Gamma^{\text{channel}} = S^{\text{channel}} \, \Gamma^{\text{initial}} \, (S^{\text{channel}})^\dagger.
\end{equation}
Next, we consider the operation of the central station. Since the operation of a 50:50 beamsplitter is
\begin{equation}
    S^{\text{BS}} \left( \frac{1}{2} \right) = \frac{1}{\sqrt{2}} \left[ 
    \begin{array}{cc}
        I_2 & I_2 \\
        -I_2 & I_2
    \end{array}
    \right],
\end{equation}
a covariance matrix of a quantum state after the 50:50 beamsplitter is
\begin{equation}
    \Gamma^{\text{50:50BS}} = \left( S^{\text{BS}}_{A'B'} \left( \frac{1}{2} \right) \right) \, \Gamma^{\text{channel}} \, \left( S^{\text{BS}}_{A'B'} \left( \frac{1}{2} \right)  \right)^\dagger,
\end{equation}
where the identity is applied to the other modes. We consider the operations of the detection efficiencies of the on-off detectors. Since detection efficiency is also modeled as a beamsplitter, this operation is as follows
\begin{equation}
\begin{split}
S^{\text{detector}} &=  S^{\text{BS}}_{AF_A} (\eta_e) \oplus S^{\text{BS}}_{A'E_A^2} (\eta_d)\\
&\quad\oplus S^{\text{BS}}_{BF_B} (\eta_e) \oplus S^{\text{BS}}_{B'E_B^2} (\eta_d)\oplus I_{E_A^1E_B^1}.  
\end{split}
\end{equation}
Then, the covariance matrix of the quantum state after this operation is
\begin{equation}
\Gamma^{\text{detector}}= S^{\text{detector}} \, \Gamma^{\text{50:50BS}}\, \left( S^{\text{detector}}  \right)^\dagger,
\end{equation}
and let $\sigma$ denote the quantum state corresponding to this covariance matrix.

We consider the detection at the central node and accept events where the detector of the mode $A'$ clicks and that of the mode $B'$ does not click. The positive operator-valued measure (POVM) corresponding to the on-off detector is 
\begin{equation}
    \{ (1-p_{d}) \ketbra{0}, \hat{I}-(1-p_{d}) \ketbra{0}\},
\end{equation}
where $\hat{I}$ is the identity operator. Then, a joint quantum state between Alice and Bob conditioned on successful heralding is
\begin{equation}
\begin{split}
    \rho_{AB} 
    &=q_1 \rho_{AB}^{(1)} - q_2 \rho_{AB}^{(2)},
\end{split}
\end{equation}
where
\begin{align}
    \rho_{AB}^{(1)} &= \tilde{\rho}_{AB}^{(1)}/\text{Tr} [\tilde{\rho}_{AB}^{(1)}],\\
    \rho_{AB}^{(2)} &= \tilde{\rho}_{AB}^{(2)}/\text{Tr} [\tilde{\rho}_{AB}^{(2)}],\\
    \tilde{\rho}_{AB}^{(1)} &= \text{Tr}_{A'E_A^1E_A^2F_AB'E_B^1E_B^2F_B} [ \sigma \ketbra{0}_{B'}   ],\\
    \tilde{\rho}_{AB}^{(2)} &= \text{Tr}_{A'E_A^1E_A^2F_AB'E_B^1E_B^2F_B} [ \sigma \ketbra{00}_{A'B'}   ],\\
    q_1 &= \frac{(1-p_{d})\text{Tr}[\sigma \ketbra{0}_{B'}]}{P},\\
    q_2 &= \frac{(1-p_{d})^2 \text{Tr}[\sigma \ketbra{00}_{A'B'}]}{P},
\end{align}
and $P$ is a probability that successful events occur:
\begin{equation}
\begin{split}
    P &= (1-p_{d}) \text{Tr} [\sigma \ketbra{0}_{B'}]\\
    &\quad - (1-p_d)^2 \text{Tr} [\sigma \ketbra{00}_{A'B'} ].
\end{split}
\end{equation}
By tracing out the modes except for $B'$, and $A'$ and $B'$ from $\Gamma^{\text{detector}}$, we obtain covariance matrices $\Gamma_{B'}$ and $\Gamma_{A'B'}$, respectively. By using these covariance matrices, we calculate the following probabilities
\begin{align}
        \text{Tr} [\sigma \ketbra{0}_{B'}] &= \frac{2}{\sqrt{\det( \Omega_2 (\Gamma_{B'} + I_2) \Omega_2^T )}},\\
        \text{Tr} [\sigma \ketbra{00}_{A'B'}] &= \frac{4}{\sqrt{\det( \Omega_4 (\Gamma_{A'B'} + I_4) \Omega_4^T )}},
\end{align}
where
\begin{equation}
    \Omega_n = \bigoplus_{i= 1}^n \omega_i =  \left[ \begin{array}{ccc}
        \omega & &  \\
         & \ddots & \\
         & & \omega
    \end{array} \right],
\end{equation}
\begin{equation}
    \omega = \left[ \begin{array}{cc}
        0 & 1 \\
        -1 & 0
    \end{array}\right].
\end{equation}

Next, we calculate the probability distribution $P(a,b|x,y)$. To this end, we first derive the covariance matrices $\Gamma^{(1)}$ and $\Gamma^{(2)}$ corresponding to $\rho_{AB}^{(1)}$ and $\rho_{AB}^{(2)}$, respectively. Let $\Gamma_{AA'BB'}$ denote the covariance matrix obtained by tracing out the environmental modes from $\Gamma^{\text{detector}}$. By appropriately reordering the modes, we obtain $\Gamma_{ABA'B'}$ which takes the following form
\begin{equation}
    \Gamma_{ABA'B'} = \left[ 
    \begin{array}{cc}
        \Gamma_{ABA'} & C \\
        C^T & \Gamma_{B'}
    \end{array}
    \right].
\end{equation}
Since the projection on vacuum can be considered as heterodyne measurement, a covariance matrix after the projection on vacuum on the mode $B'$ is
\begin{equation}
\begin{split}
        \Gamma'_{ABA'} &= \Gamma_{ABA'}- C (\Gamma_{B'} + I_2)^{-1} C^T\\
        &= \left[ 
        \begin{array}{cc}
            \Gamma'_{AB} & C' \\
            C'^T & \Gamma'_{A'}
        \end{array}
        \right].
\end{split}
\end{equation}
Therefore, $\Gamma^{(1)}$ and $\Gamma^{(2)}$ can be derived as follows
\begin{align}
        \Gamma^{(1)} &= \text{Tr}_{A'} \left[ \Gamma_{ABA'} \right] = \Gamma'_{AB},\\
        \Gamma^{(2)} &= \Gamma'_{AB} - C' (\Gamma_{A'}+I_2)^{-1} C'^T.
\end{align}
Let $\alpha$ and $\beta$ be displacement amplitudes of Alice and Bob, respectively. Then, after performing these displacement operations, a displacement vector of the joint quantum state between Alice and Bob becomes
\begin{equation}
    d_{AB} = 2\left[ \text{Re} (\alpha), \text{Im} (\alpha), \text{Re} (\beta), \text{Im} (\beta) \right]^T.
\end{equation}
First, we calculate a probability $P(0,0|x,y)$ where both of Alice's and Bob's on-off detectors do not click. POVM of these detectors can be expressed as
\begin{equation}
    \{ (1-p_{d}^e) \ketbra{0}{0} , \hat{I} - (1-p_{d}^e) \ketbra{0}{0}  \}.
\end{equation}
Then, we calculate this probability in the following way
\begin{equation}
\begin{split}
    P(0,0|x,y) &= \text{Tr} \left[ \rho_{AB} (1-p_{d}^e)^2 \ketbra{00}{00}_{AB}  \right] \\
    &=(1-p_{d}^e)^2  \text{Tr} \left[q_1 \rho_{AB}^{(1)} \ketbra{00}{00}_{AB}  \right]\\
    &\quad -(1-p_{d}^e)^2 \text{Tr} \left[q_2 \rho_{AB}^{(2)} \ketbra{00}{00}_{AB}  \right].
\end{split}
\end{equation}
Here, let $\chi^{(1)}(x)$ be a characteristic function of $\rho^{(1)}_{AB}$ and $\chi_0(x)$ be a characteristic function of projection onto vacuum defined as follows
\begin{align}
        \chi^{(1)} (x) &= \exp (-\frac{1}{2} x^T (\Omega_4 \Gamma^{(1)} \Omega_4^T) x -i (\Omega_4 d_{AB})^T x), \\
        \chi_0 (x) &= \exp(-\frac{1}{2} x^T (\Omega_4 I_4 \Omega_4^T) x),
\end{align}
where $x \in \mathbb{R}^4$ is a real vector. Then, the following holds
\begin{widetext}
\begin{equation}
    \begin{split}
        \text{Tr} \left[q_1 \rho_{AB}^{(1)} \ketbra{00}{00}_{AB}   \right] &= \frac{q_1}{\pi^2} \int \chi^{(1)} (x) \chi_0 (x) \, dx \\
        &= \frac{4q_1}{\sqrt{\det (\Omega_4 (\Gamma^{(1)} + I_4)) \Omega_4^T} } \exp (-\frac{1}{2} (\Omega_4 d_{AB})^T (\Omega_4 (\Gamma^{(1)} + I_4) \Omega_4^T)^{-1} (\Omega_4 d_{AB})) \\
    \end{split}
\end{equation}
\end{widetext}
We can analogously calculate the term $\text{Tr} \left[q_2 \rho_{AB}^{(2)} \ketbra{00}{00}_{AB}  \right]$ which allows us to derive the probability $P(0,0|x,y)$. The marginal probabilities that Alice's and Bob's on-off detectors do not click, denoted by $P_A(0|x)$ and $P_B(0|y)$, respectively, are calculated in the same manner. By using these probabilities, we calculate the remaining probabilities as follows
\begin{align}
    P(0,1|x,y) &= P_A(0|x) - P(0,0|x,y), \\
    P(1,0|x,y) &= P_B(0|y) - P(0,0|x,y),\\
    P(1,1|x,y) &= 1 - P_A(0|x) - P_B(0|y) + P(0,0|x,y).
\end{align}

Next, we describe the procedure for Protocol B. In this protocol, Bob prepares a two-mode squeezed state and applies a beamsplitter with transmissivity $\tau$ to one of the two modes. Let $B, B'$ and $C$ denote the transmitted mode, the reflected mode, and the remaining mode of the two-mode squeezed state, respectively. The covariance matrix of the resulting quantum state before measurement can be calculated in the same way as in Protocol A. Let $\sigma$ denote this quantum state. Then, an unnormalized quantum state shared by Alice and Bob is expressed as 
\begin{equation}
\begin{split}
    \tilde{\rho}_{AB} &= \text{Tr}_{/AB} [ \sigma (\hat{I}-(1-p_{d})\ketbra{0}{0})_{A'} (1-p_{d})\ketbra{0}{0}_{B'}  \\
    &\quad \otimes (\hat{I}-(1-p_{d})\ketbra{0}{0})_{C} ]\\
    &= (1-p_{d}) \text{Tr}_{/AB} [ \sigma \ketbra{0}{0}_{B'}   ] \\
    &\quad - (1-p_{d})^2 \text{Tr}_{/AB} [ \sigma \ketbra{00}{00}_{A'B'}  ]\\
    &\quad - (1-p_{d})^2 \text{Tr}_{/AB} [ \sigma \ketbra{00}{00}_{B'C'} ]\\
    &\quad + (1-p_{d})^3 \text{Tr}_{/AB} [\sigma \ketbra{000}{000}_{A'B'C}  ],
\end{split}
\end{equation}
where $/AB$ denotes all the modes except for $A$ and $B$. By appropriately normalize this quantum state, we obtain the joint quantum state between Alice and Bob which takes the following form
\begin{equation}
    \rho_{AB}   = q_1 \rho_{AB}^{(1)} - q_2 \rho_{AB}^{(2)} - q_3 \rho_{AB}^{(3)} + q_4 \rho_{AB}^{(4)}.
\end{equation}
By using the same approach as Protocol A, we calculate the coefficients $q_1, \ldots, q_4$ and covariance matrices corresponding to $\rho_{AB}^{(1)}, \ldots, \rho_{AB}^{(4)}$. Then, we derive the probability distribution $P(a,b|x,y)$ in the same way as Protocol A.

\section{KEY RATE}\label{sec:Keyrate}
In this section, we describe how to calculate key rates of our DI-QKD protocols. We focus on the asymptotic key rates. The key rate $K$ of our protocols can be expressed as~\cite{Devetak2005}
\begin{equation}\label{eq:keyrate}
    K = P(H(A|x^*, E) - H(A|B,x^*,y^*)),
\end{equation}
where $A$, $B$, and $E$ represent Alice's, Bob's and Eve's systems, respectively, $x^*$ ($y^*$) denotes Alice's (Bob's) input for the key generation rounds, i.e., $x^* = 0$ ($y^* = 2$), $P$ is the probability that the successful events occur, $H(A|x^*, E)$ is the conditional von Neumann entropy between Alice and Eve, and $H(A|B, x^*, y^*)$ represents the error correction cost between Alice and Bob. The first term in (\ref{eq:keyrate}) is non-trivial to compute because it involves Eve's system, which is uncharacterized. Instead, we compute a lower bound on this quantity using the numerical optimization framework developed in Refs.~\cite{Navascues2008, Masanes2011, NietoSilleras2014, Bancal2014, Brown2024}. Let $\{M_{a|x}^A \}$ and $\{ M_{b|y}^B  \}$ denote the measurement operators for Alice and Bob. Also, let $\{M_{a|x^*}^A \}$ denote the measurement operators for the key generation rounds where $x^* =0$. For $a \in \{ 0,1\}$, we define $\hat{M}_a = (1-p_n)M_{a|x^*}^A + p_n M_{a \oplus 1 |x^*}^A$ where $p_n$ corresponds to the probability that Alice flips her key bits for the noisy preprocessing. Let $m \in \mathbb{N}$, and let $t_1, \ldots, t_m$ and $w_1, \ldots, w_m$ be the nodes and weights of an $m$-point Gauss-Radau quadrature on $[0,1]$ with an endpoint $t_m =1$, respectively. Furthermore, denote $\alpha_i = \frac{3}{2} \max \{ \frac{1}{t_i} , \frac{1}{1-t_i} \}$. Then, we can obtain a lower bound on the conditional entropy $H(A|x^*, E)$ by solving the following optimization problem~\cite{Brown2024}
\begin{widetext}
\begin{equation}
\label{eq:optimization}
    \begin{split}
        c_m + \inf \quad &\sum_{i=1}^{m-1} \frac{w_i}{t_i \ln 2 } \sum_{a=0}^1 \mel{\psi}{\hat{M}_a (Z_{a,i} + Z_{a,i}^* +(1-t_i) Z_{a,i}^* Z_{a,i}) + t_i Z_{a,i} Z_{a,i}^*}{\psi}\\
        \text{subject to} \quad
        &\mel{\psi}{M_{a|x}^A M_{b|y}^B}{\psi} = P(a,b|x,y) \qquad \text{for all} ~ a, b, x, y\\
        &\sum_a M_{a|x}^A = \sum_{b} M_{b|y}^B = I \qquad \text{for all} ~ x, y\\
        &M_{a|x}^A \geq 0, \quad M_{b|y}^B \geq 0 \qquad \text{for all} ~ a, b, x, y\\
        & Z_{a,i}^* Z_{a,i} \leq \alpha_i^2, \quad Z_{a,i} Z_{a,i}^* \leq \alpha_i^2 \qquad \text{for all} ~ a, i\\
        &\left[M_{a|x}^A, M_{b|y}^B \right] = \left[M_{b|y}^A, Z_{a,i}^{(*)} \right] = \left[ M_{a|x}^B, Z_{a,i}^{(*)} \right] =0 \qquad \text{for all} ~ a, b, x, y, i\\
        &M_{a|x}^A, M_{b|y}^B, Z_{a,i} \in B(\mathcal{H}) \qquad \text{for all} ~ a, b, x, y, i
    \end{split}
\end{equation}
\end{widetext}
where the infimum is taken over all quantum states $\ket{\psi}$ and bounded operators $M_{a|x}^A$, $M_{b|y}^B$ and $Z_{a, i}$, $c_m = \frac{2p_n (1-p_n)}{m^2 \ln 2} + \sum_{i=1}^{m-1} \frac{w_i}{t_i \ln2}$, and $B(\mathcal{H})$ is the set of bounded operators on a Hilbert space $\mathcal{H}$. 

Since no efficient direct method exists for solving this optimization problem, we relax it into semidefinite programming (SDP) using the Navascu\'{e}s-Pironio-Ac\'{i}n (NPA) hierarchy \cite{Navascues2008}. Here, we give an overview of the NPA hierarchy. Let us define a set of Alice's measurement operators $\mathcal{A} = \{M_{a|x}^A  \}$ on a Hilbert space $\mathcal{H}$. Also, in a similar way, we define a set of Bob's and Eve's measurement operators $\mathcal{B}$ and $\mathcal{Z}$, respectively. Let $\mathcal{E}$ denote a set of these operators and the identity $\hat{I}$, that is, $\mathcal{E} = \hat{I} \cup \mathcal{A} \cup \mathcal{B} \cup \mathcal{Z}$. Suppose $O_i \, (i \in \{1, \ldots, m \})$ denotes a linear combination of products of operators in $\mathcal{E}$ and we minimize (or maximize) a linear combination of those expectation values $\ev{O_i^\dagger O_j} = \text{Tr} \left[ \ketbra{\psi}{\psi} O_i^\dagger O_j \right]$ where $\ket{\psi}$ is a quantum state on $\mathcal{H}$. The objective function in (\ref{eq:optimization}) is also expressed in this form. To relax this problem into SDP, we construct a $k$-th level Gram matrix $\gamma^{(k)}$. Entries of $\gamma^{(k)}$ are expressed as $\gamma_{ij}^{(k)} = \ev{(O_i^k)^\dagger O_j^k}$, where $O_i^k$ denotes a product of at most $k$ operators in $\mathcal{E}$. Then, the objective function is expressed as a linear combination of entries of the Gram matrix, and the Gram matrix is positive semidefinite. Therefore, the optimization problem can be interpreted as SDP over the Gram matrix. There are some available tools for performing the NPA hierarchy and for solving SDP. We perform the NPA hierarchy using NCPOL2SDPA~\cite{Wittek2015} and solve SDP using MOSEK~\cite{Mosek2024}.

To calculate the second term in (\ref{eq:keyrate}), we include the effect of the noisy preprocessing into the probability distribution $P(a,b|x^*,y^*)$. After the noisy preprocessing, the probability distribution comes to be
\begin{equation}
\begin{split}
    P'(a,b|x^*,y^*) &= (1-p_n) P(a, b|x^*, y^*) \\
    &\quad + p_n P(a\oplus1, b|x^*, y^*)
\end{split}
\end{equation}
Then, the conditional entropy between Alice and Bob can be calculated by
\begin{equation}
\begin{split}
    H(A&|B, x^*, y^*) = \sum_{a, b} H(P'(a,b|x^*, y^*))\\
    &\quad - \sum_{b} H(P'(0,b|x^*, y^*) + P'(1,b|x^*, y^*)),
\end{split}
\end{equation}
where $H(p) \equiv -p \log_2 p$.

\section{RESULTS AND DISCUSSION}\label{sec:results}
In this section, we present and discuss the results of our key rate calculations. We assume a symmetric setup in which the central station is located midway between Alice and Bob, i.e., $\eta_A = \eta_B = \eta$. Channel loss is modeled as 0.2 dB/km corresponding to $\eta = 10^{-0.02 L}$ where $L$ is the distance in kilometers from each party to the central station. For simplicity, we assume that the on-off detectors at the central station and those of Alice and Bob have the same dark count probabilities; $p_d^e = p_d$, and we assume the mean photon number of Alice's and Bob's two-mode squeezed states are the same; $\bar{n}_A = \bar{n}_B = \bar{n}$. For all results, we numerically optimize Alice's and Bob's displacement operations and the probability $p_n$. We fix the number of the nodes in the Gauss-Radau quadrature to $m = 8$.

\begin{figure}[t]
 \centering
 \includegraphics[keepaspectratio, scale=0.5]{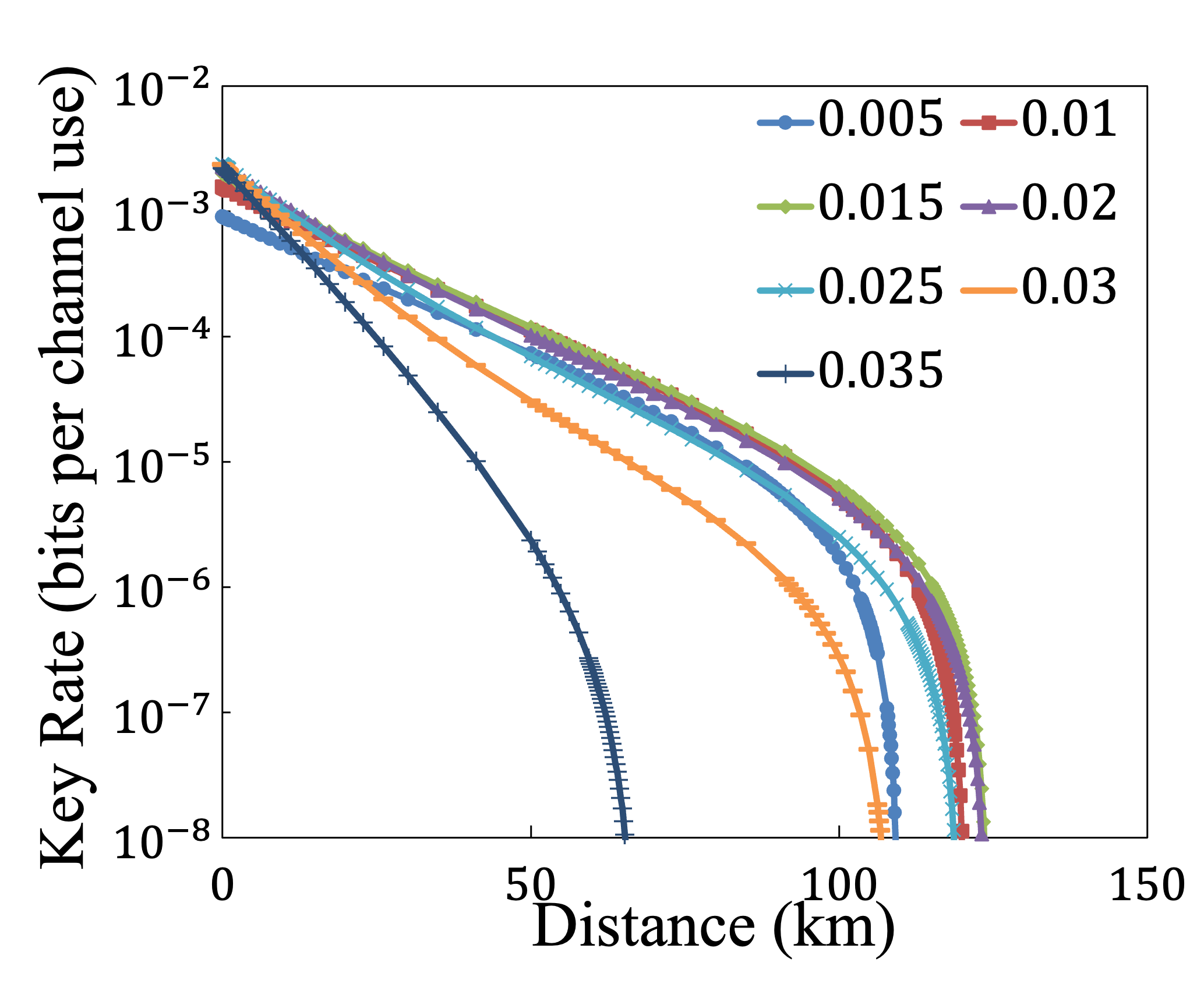}
 \caption{Key rate of Protocol A versus distance $L$ for different mean photon number $\bar{n}$. We take $\eta_e = \eta_d = 0.95$ and $p_d = 10^{-6}$.}
 \label{fig:DifferentV}
\end{figure}

Fig.~\ref{fig:DifferentV} shows the key rates of Protocol A as a function of distance for various values of the mean photon number of two-mode squeezed states $\bar{n}$. We fix the parameters $\eta_d = 0.95$, $\eta_e = 0.95$, and $p_d = 10^{-6}$. We observe that the key rate exhibits an optimal mean photon number which maximizes the achievable communication distance. The trade-off can be understood as follows. When the mean photon number is large, the probability that photons are emitted and reach the central station increases, resulting in a higher success probability $P$. However, the contribution from multiphoton components also grows, which weakens the Bell violation and increases vulnerability to loopholes. On the other hand, when the mean photon number is small, the effect of such multiphoton events is suppressed, but the success probability becomes very low. In this case, dark counts in the central station detectors become more likely to dominate, as they are indistinguishable from legitimate clicks. This degrades the key rate and may eliminate the possibility of secure key generation. Thus, an intermediate value of $\bar{n}$ optimizes performance. A similar behavior is observed for Protocol B.

\begin{figure}[htbp]
 \centering
 \includegraphics[keepaspectratio, scale=0.5]{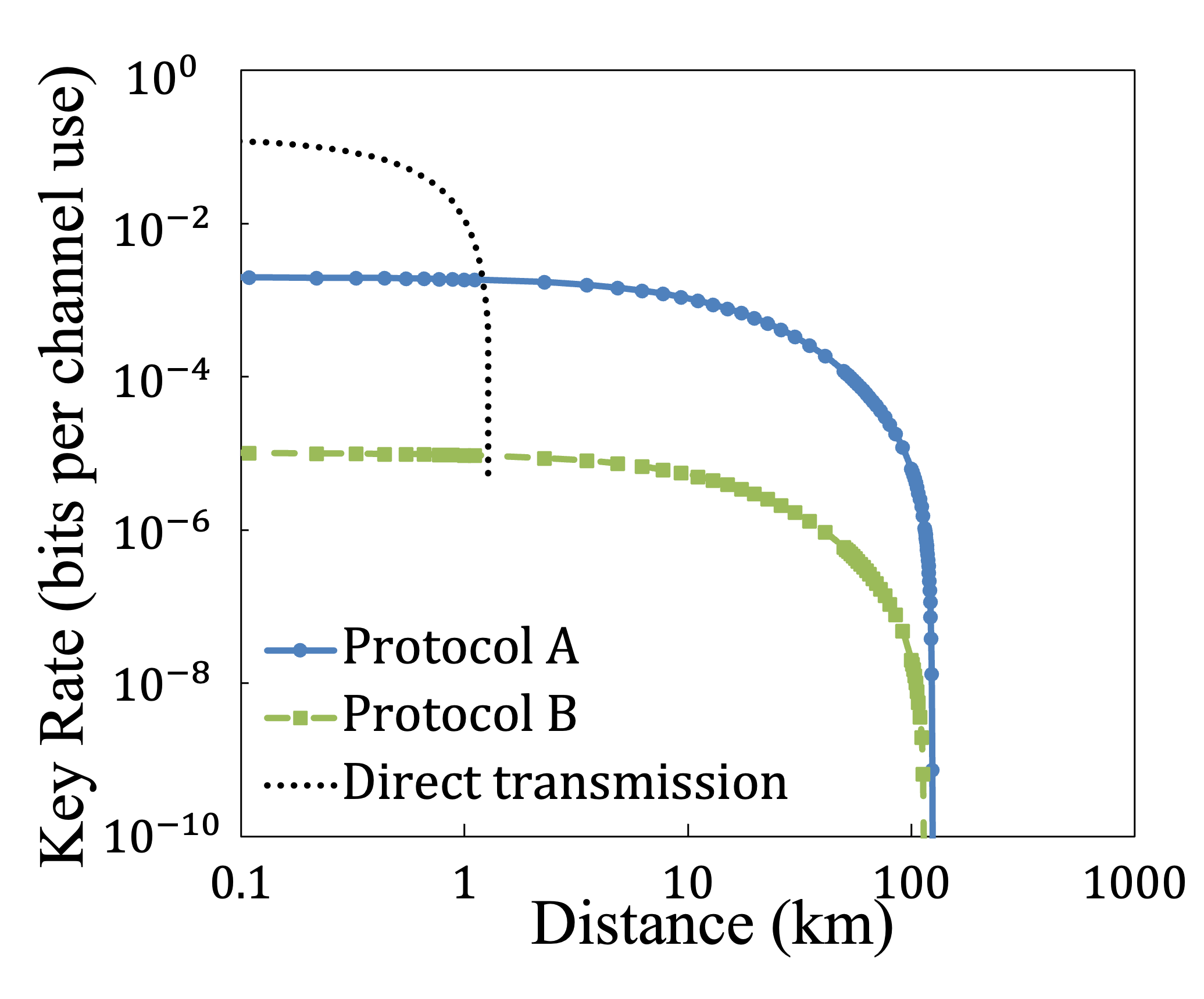}
 \caption{Key rate versus distance $L$. Blue circles, green squares and black dashed line represent key rates of Protocol A,  Protocol B and the direct transmission protocol, respectively. We take $\eta_d = \eta_e = 0.95$ and $p_d = 10^{-6}$. We use $\bar{n} = 0.015$ for Protocol A, $\bar{n} = 0.01$ for Protocol B, and we optimize the mean photon number for each key rate for the direct transmission protocol.}
 \label{fig:Comparison}
\end{figure}

Fig.~\ref{fig:Comparison} compares the performance of our protocols with a direct transmission protocol where Alice and Bob directly perform the displacement-based measurements on a shared two-mode squeezed state (see Appendix~\ref{appendix:directtransmission}). The circles represent key rates of Protocol A, the squares represent key rates of Protocol B, and the black dotted line represents key rates of the direct transmission protocol where $\eta_d = \eta_e = 0.95$ and $p_d = 10^{-6}$. We take $\bar{n} = 0.015$ for Protocol A and $\bar{n} = 0.01$ for Protocol B. The direct transmission protocol is optimized over $\bar{n}$, and we also optimize the beamsplitter transmissivity $\tau$ for Protocol B. It can be observed that both of our protocols outperform the direct transmission approach by up to two orders of magnitude in distance, enabled by the use of single-photon interference. While our protocols can reach longer distances than the direct transmission, the key rates of our protocols at short distances are smaller than those of the direct transmission because of the overhead of the single-photon interference requirements. Also, the key rates of Protocol A are consistently larger than those of Protocol B. This is attributed to the lower success probability of Protocol B, which requires a simultaneous click in both the heralding and central station detectors.

\begin{figure}[htbp]
 \centering
 \includegraphics[keepaspectratio, scale=0.5]{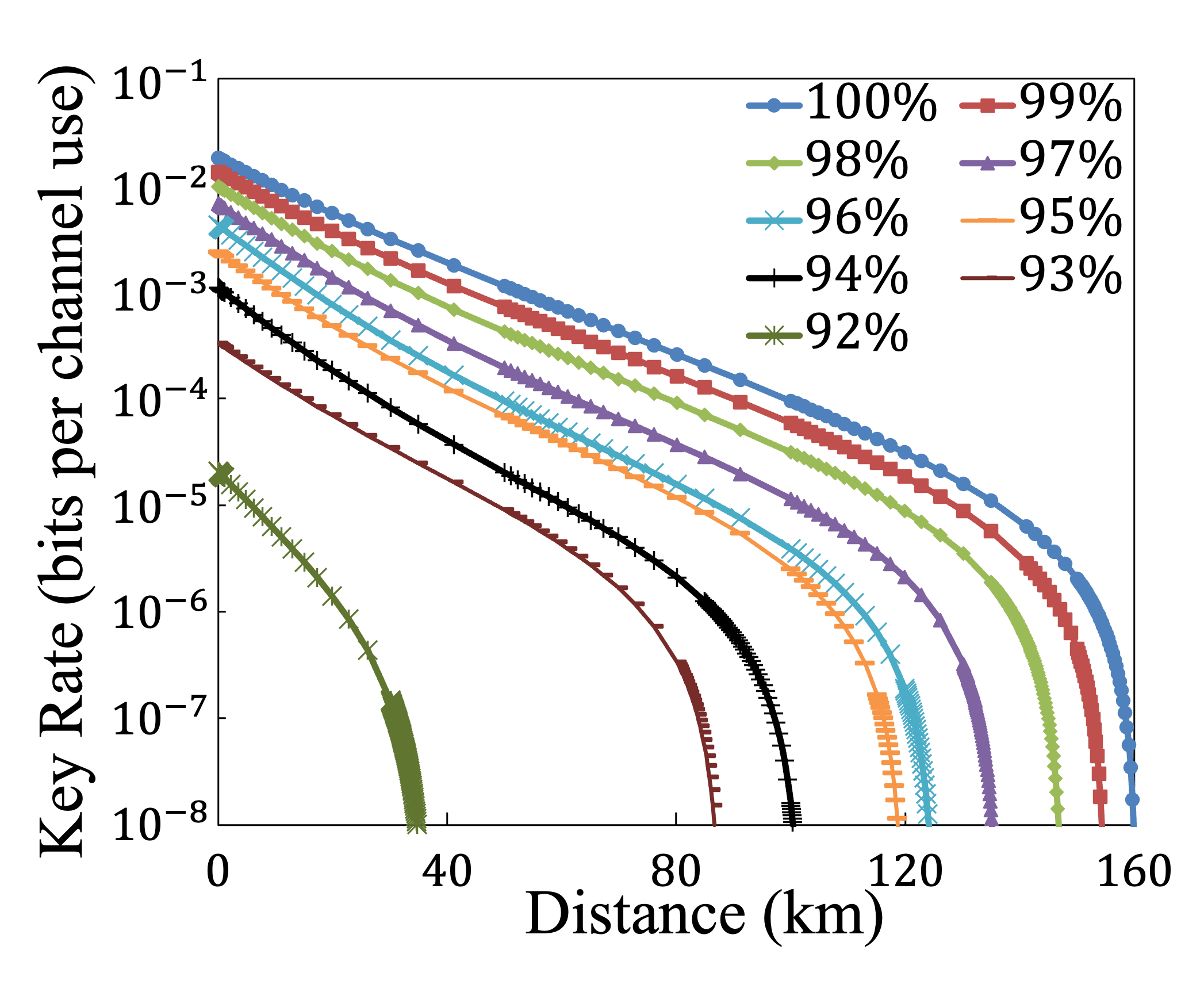}
 \caption{Key rate of protocol A versus distance $L$ for different detection efficiency $\eta_e$. We take $\eta_d = 0.95$ and $p_d = 10^{-6}$.}
 \label{fig:DifferentEtaeA}
\end{figure}

\begin{figure}[htbp]
 \centering
 \includegraphics[keepaspectratio, scale=0.5]{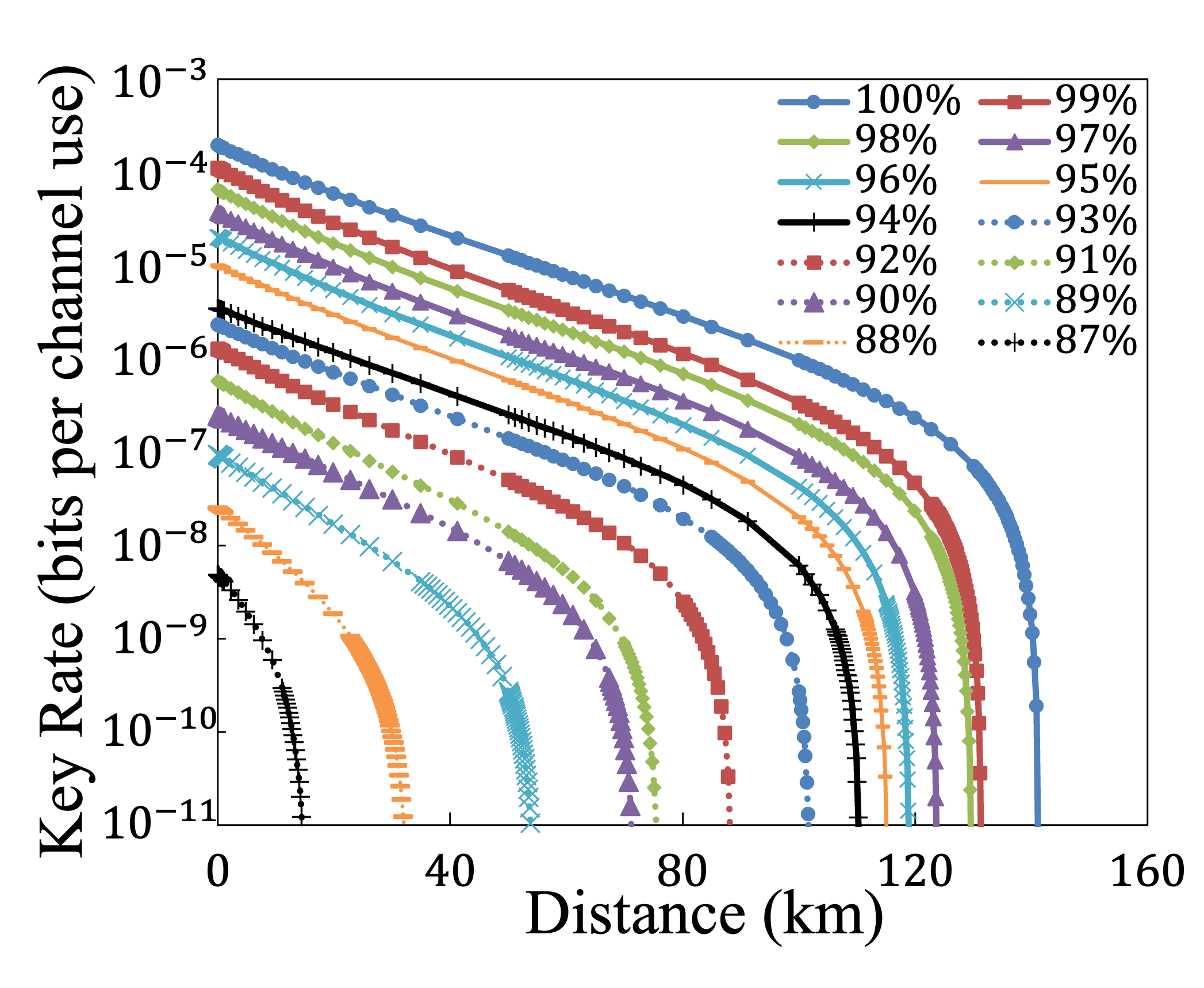}
 \caption{Key rate of protocol B versus distance $L$ for different detection efficiency $\eta_e$. We take $\eta_d = 0.95$ and $p_d = 10^{-6}$.}
 \label{fig:DifferentEtaeB}
\end{figure}

\begin{figure}[htbp]
 \centering
 \includegraphics[keepaspectratio, scale=0.5]{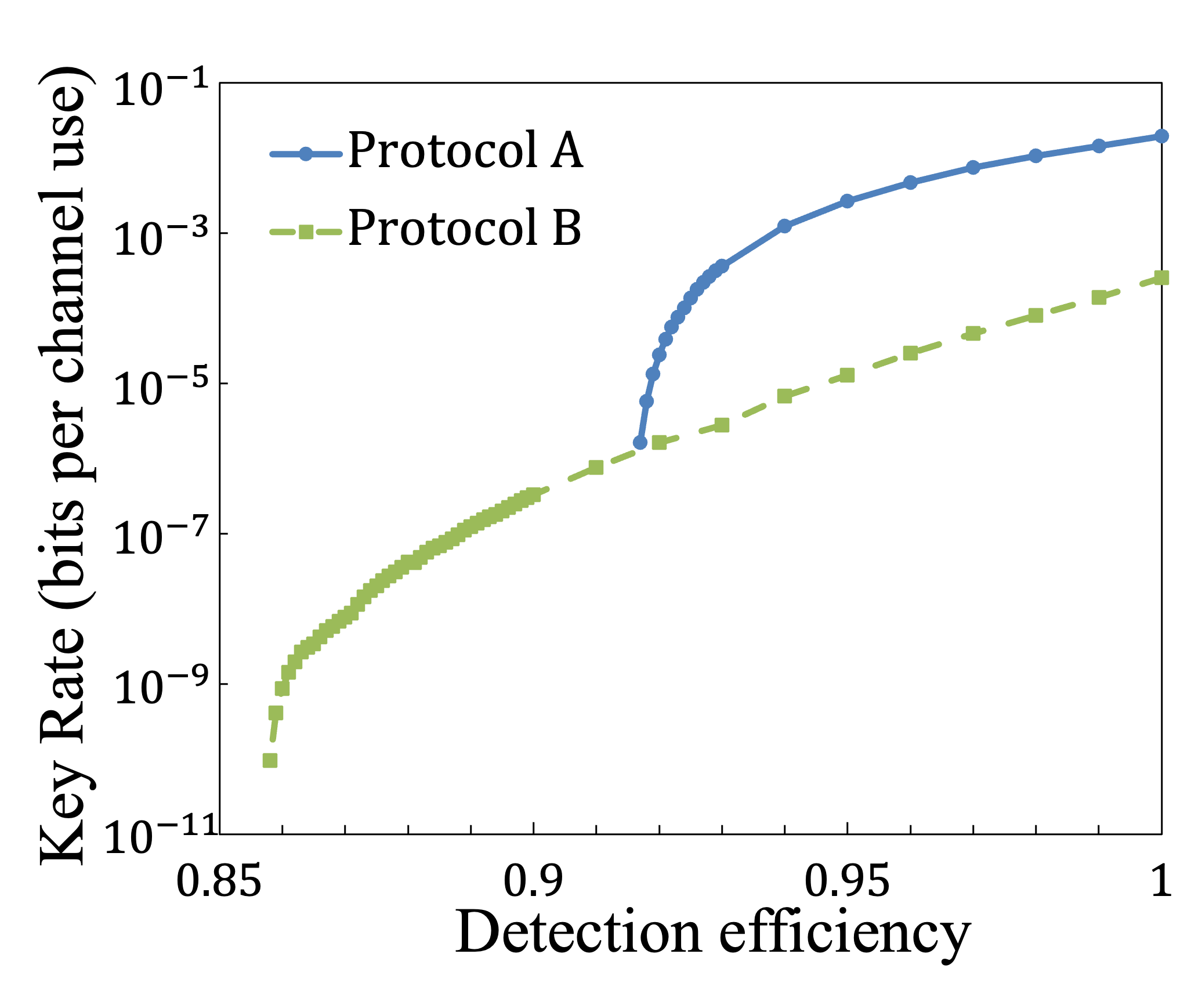}
 \caption{Key rate versus detection efficiency $\eta_e$, where we optimize the mean photon number $\bar{n}$ for each point. Blue circles and green circles show key rates of protocol A and protocol B, respectively. We take $\eta_d = 1$ and $p_d = 0$.}
 \label{fig:DetectionEff}
\end{figure}

In Fig.~\ref{fig:DifferentEtaeA} and Fig.~\ref{fig:DifferentEtaeB}, we examine the dependence of the key rate on the detection efficiency $\eta_e$ for Protocols A and B, respectively. Here, we use $\eta_d = 0.95$ and $p_d = 10^{-6}$, and for each value of $\eta_e$, we optimize both the mean photon number of the two-mode squeezed states and the transmissivity $\tau$ for Protocol B. We find that detection efficiency $\eta_e$ has significant impact on key rates of our protocols. While channel loss can be effectively mitigated via heralded entanglement generation, the effect of local detection inefficiency cannot be compensated. Therefore, high detection efficiency is essential to obtain loophole-free violations of Bell inequalities. We also present the key rates of our protocols against detection efficiency $\eta_e$ with ideal parameters, i.e., $\eta = \eta_d = 1$ and $p_d = 0$ in Fig.~\ref{fig:DetectionEff}, again optimizing the mean photon number of two-mode squeezed states $\bar{n}$ for each point. We find that Protocol A requires a minimum detection efficiency of 0.917 to produce a positive key rate while Protocol B achieves it for an efficiency as low as 0.858. This difference is attributed to the distinct quantum states shared between Alice and Bob in the protocols. In Protocol A, the post-selected state is approximately a single-photon Bell state:
\begin{equation}
    \frac{1}{\sqrt{2}} (\ket{01} + \ket{10}),
\end{equation}
while in Protocol B, the shared state is approximately:
\begin{equation}
    \frac{1}{\sqrt{1+r^2}}( \ket{00} + r \ket{11}),
\end{equation}
where $r$ is a parameter which can be tuned by changing the transmissivity $\tau$. By appropriately choosing $r$, the state in Protocol B exhibits the anomaly of nonlocality~\cite{Eberhard1993, Methot2007}, which reduces the required detection efficiency for violating a Bell inequality. This anomaly leads to a lower threshold efficiency for Protocol B compared to Protocol A (see Appendix~\ref{appendix:anomaly}). These values are achievable with current state-of-the-art~\cite{Liu2022, Reddy2020, Chang2021}. Therefore, our protocols which consist only standard quantum optics tools can be experimentally demonstrated with current technology.

\section{CONCLUSION}\label{sec:conclusion}
In this work, we have proposed two practical long-distance DI-QKD protocols that leverage a heralding scheme based on single-photon interference. Both protocols rely exclusively on standard quantum optics components, including two-mode squeezed states, displacement operations, and on-off detectors--thus ensuring compatibility with current experimental capabilities. Using a numerical optimization framework, we have calculated the asymptotic key rates for both protocols and shown that they significantly outperform non-heralded DI-QKD protocols in terms of communication distances. In particular, our analysis reveals an optimal mean photon number for maximizing secure communication range, and we identify critical thresholds for detector efficiency under realistic imperfections.

Finally, we provide possible directions for future work. First, while we analyze the asymptotic key rates in this paper, it is important future work to analyze performance of our protocols with finite-size security. Such analysis can be realized by using the entropy accumulation theorem~\cite{ArnonFriedman2018, Dupuis2020} or the generalized entropy accumulation theorem~\cite{Metger2022}. 
Moreover, although our required detection efficiencies are compatible with state-of-the-art technology, further relaxation of these requirements would improve the practicality of experimental implementation. One promising way to achieving this is the use of local Bell tests~\cite{Ernest2024, RoyDeloison2025}. In lieu of the standard two-party setup, Bob is able to route his quantum state with a switch at random to another device that is located near the source, $\hat{T}$. This allows Bob to perform a Bell test between Alice's device $A$ and Bob's device near the source $\hat{T}$, or between Alice's device $A$ and Bob's device farther away from the source $B$. The Bell test configuration of $A\hat{T}$ ensures Alice's (near) honest behavior in the Bell test configuration of $AB$, by limiting the possible set of states and measurements of Alice in the $AB$ configuration, thus allowing a reduction in the detection efficiency thresholds, shown in Refs.~\cite{Ernest2024, RoyDeloison2025}. 

{\it Note added:} After completing this work, we became aware of Ref.~\cite{Moradi2025} which proposes protocols similar to our Protocols A and B and analyzes their performance in the finite block-length regime. In Ref.~\cite{Moradi2025}, an ideal single-photon Bell state is assumed. Also, the maximum number of photons coming to the central station is assumed to be one, which effectively makes the Bell measurement to be photon number resolving. In contrast, we employ on-off detectors for all measurements in two protocols, resulting the heralded single-photon Bell state nonideal and no photon number resolving capability for the Bell measurement. These assumptions cause small performance gaps between them, e.g., the minimum required detection efficiency in the asymptotic limit.  


\begin{acknowledgments}
We thank Yazeed K. Alwehaibi, Raj B. Patel, and Magdalena Stobi{\'{n}}ska for helpful discussions. This work is supported by JST SPRING, Grant No. JPMJSP2123, JST Moonshot R\&D, Grant No. JPMJMS226C and Grant No. JPMJMS2061, JST CRONOS, Grant No. JPMJCS24N6, JST ASPIRE, Grant No. JPMJAP2427, and JST COI-NEXT, Grant No. JPMJPF2221. We also acknowledge support from the Danish National Research Foundation, Center for Macroscopic Quantum States (bigQ, DNRF0142), the  European Union’s Horizon Europe research and innovation programme under the project ``Quantum Security Networks Partnership'' (QSNP, grant agreement no. 101114043), and from Innovation Fund Denmark (CyberQ, grant agreement no. 3200-00035B).
\end{acknowledgments}

\appendix
\section{Direct transmission protocol}\label{appendix:directtransmission}
\begin{figure}[htbp]
 \centering
 \includegraphics[keepaspectratio, scale=0.45]{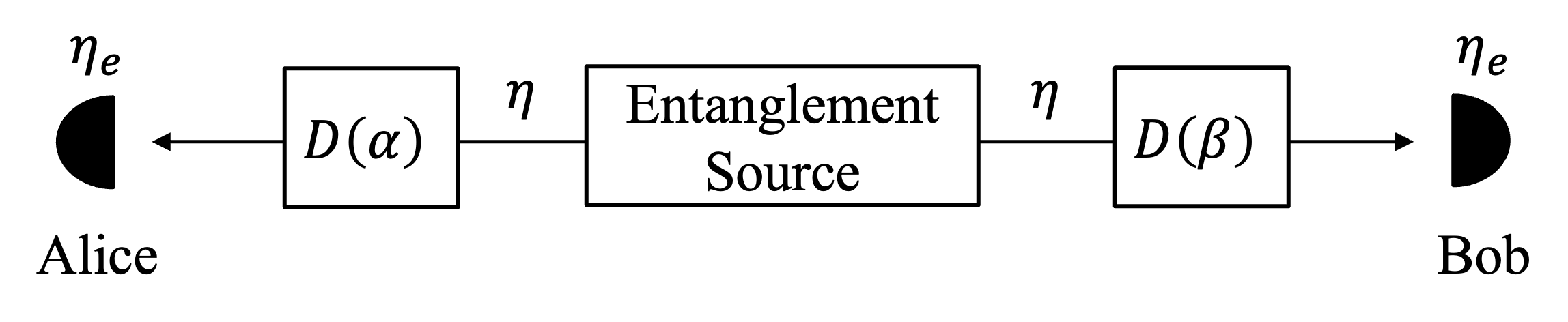}
 \caption{Schematic of a DI-QKD protocol without the single-photon interference. An entangled state is generated at an entanglement source which is in the middle between Alice and Bob. Alice and Bob perform the displacement-based measurements on the entangled state. They share a secret key in the same approach as our protocols with single-photon interference. When the entanglement source generates a two-mode squeezed state, this protocol is equivalent to the direct transmission protocol discussed in Appendix~\ref{appendix:directtransmission}. The entanglement source generates Bell states in the protocol which we consider in Appendix~\ref{appendix:anomaly}.
}
 \label{fig:Directsetup}
\end{figure}

We describe the direct transmission protocol which is schematically shown in Fig.~\ref{fig:Directsetup}. In this protocol, Alice and Bob directly perform the displacement-based measurements on a two-mode squeezed state generated in the middle between Alice and Bob and sent through pure-loss channels with transmissivity $\eta$. Let $\eta_e$ and $p_d$ denote detection efficiency and dark count probability of Alice's and Bob's on-off detectors. Alice (Bob) randomly selects measurement inputs $x \in \{0, 1 \}$ ($y \in \{0,1,2  \}$) corresponding to the displacement $\alpha$ ($\beta$) and obtains measurement outputs $a \in \{0 ,1  \}$ ($\beta \in \{0,1\}$) corresponding to click and no-click of the on-off detectors. Part of the events with $x = 0$ and $y = 2$ is used to generate a secret key and the rest of the events are used for parameter estimation. Furthermore, Alice and Bob perform the noisy preprocessing on the key generation rounds.

We derive a joint probability distribution $P(a, b|x, y)$ of Alice and Bob for this protocol. A covariance matrix of an initial quantum state is
\begin{equation}
    \Gamma^{\text{initial}} = \Gamma_{AB}^{\text{TMSV}} (\bar{n}) \oplus I_{E_AE_B},
\end{equation}
where $E_A$ and $E_B$ are environmental modes corresponding to losses and $\bar{n}$ is the mean photon number of the two-mode squeezed state. As our protocols, we effectively shift the detection efficiency of Alice's and Bob's on-off detectors prior to their displacement operations. We combine two beamsplitters corresponding to the channel loss $\eta$ and the detection efficiency $\eta_e$ as one beamsplitter with transmissivity $\eta \eta_e$. Then, the operation of this beamsplitter is expressed as follows:
\begin{equation}
    S^{\text{channel}} = S^{\text{BS}}_{AE_A} (\eta \eta_e) \oplus S^{\text{BS}}_{BE_B} (\eta \eta_e).
\end{equation}
Then, a covariance matrix of a quantum state after channel transmission is
\begin{equation}
    \Gamma^{\text{channel}} = S^{\text{channel}} \, \Gamma^{\text{initial}} \, (S^{\text{channel}})^\dagger.
\end{equation}
By tracing out the environmental modes, we obtain a joint quantum state between Alice and Bob $\rho_{AB}$ whose covariance matrix $\Gamma_{AB}$ is expressed as follows
\begin{equation}
    \Gamma_{AB} = \text{Tr}_{E_AE_B} [\Gamma^{\text{channel}}].
\end{equation}
A displacement vector of this state after the displacement operations is
\begin{equation}
    d_{AB} = 2\left[ \text{Re} (\alpha), \text{Im} (\alpha), \text{Re} (\beta), \text{Im} (\beta) \right]^T.
\end{equation}
Let $\Gamma_A$, $\Gamma_B$, $d_A$ and $d_B$ be defined as follows
\begin{align}
        \Gamma_A &= \text{Tr}_B [\Gamma_{AB}], \\
        \Gamma_B &= \text{Tr}_A [\Gamma_{AB}], \\
        d_A &= \text{Tr}_B [d_{AB}], \\
        d_B &= \text{Tr}_A [d_{AB}].
\end{align}
Then, we can calculate a probability that both of Alice's and Bob's on-off detectors do not click $P(0,0|x,y)$
\begin{widetext}
\begin{equation}
    \begin{split}
        P(0,0|x,y) &= \frac{4(1-p_{d})^2}{\sqrt{\det (\Omega_4 (\Gamma_{AB} + I_4)) \Omega_4^T} } \exp (-\frac{1}{2} (\Omega_4 d_{AB})^T (\Omega_4 (\Gamma_{AB} + I_4) \Omega_4^T)^{-1} (\Omega_4 d_{AB})).
    \end{split}
\end{equation}
Also, we can calculate a probability that Alice's on-off detector does not click
\begin{equation}
    \begin{split}
        P_A(0|x) &= \frac{2(1-p_{d})}{\sqrt{\det(\Omega_2 (\Gamma_A + I_2) \Omega_2^T)}} \exp (-\frac{1}{2} (\Omega_2 d_A)^T (\Omega_2 (\Gamma_A + I_2 )\Omega_2^T)^{-1} (\Omega_2 d_A)),
    \end{split}
\end{equation}
and the probability that Bob's on-off detector does not click
\begin{equation}
    \begin{split}
        P_B(0|y) &= \frac{2(1-p_{d})}{\sqrt{\det(\Omega_2 (\Gamma_B + I_2) \Omega_2^T)}} \exp (-\frac{1}{2} (\Omega_2 d_B)^T (\Omega_2 (\Gamma_B + I_2 )\Omega_2^T)^{-1} (\Omega_2 d_B)),
    \end{split}
\end{equation}
The remaining probabilities are calculated as follows
\begin{equation}
    \begin{split}
        P(0, 1|x, y) &= P_A(0|x) - P(0, 0|x, y), \\
        P(1, 0|x, y) &= P_B(0|y) - P(0, 0|x, y), \\
        P(1, 1|x, y) &= 1 - P_A(0,x) - P_B(0|y) + P(0, 0|x, y).
    \end{split}
\end{equation}

The key rate $K$ of this protocol is
\begin{equation}
    K = H(A|E, x^*) - H(A|B, x^*, y^*),
\end{equation}
where $x^*$ and $y^*$ denote Alice's and Bob's inputs for the key generation rounds, respectively. We calculate key rates of the direct transmission protocol in the same way as our protocols.

\end{widetext}

\section{Detection efficiency relaxation with anomaly of non-locality}\label{appendix:anomaly}
Here, we analyze the effect of the anomaly of non-locality in terms of required detection efficiency in DI-QKD protocols. We consider the following setup, shown in Fig.~\ref{fig:Directsetup}. A Bell state is generated at the middle of Alice and Bob, and the parties perform the displacement-based measurements on the quantum state. The detection efficiencies of Alice's and Bob's on-off detectors are $\eta_e$. We do not consider losses of channel transmission. This protocol is equivalent to the direct transmission protocol with $\eta = 1$ except for the generated quantum state.

Let the generated Bell state be one of the following states,
\begin{equation}\label{eq:0110}
    \frac{1}{\sqrt{1+r^2}} (\ket{01} + r \ket{10}),
\end{equation}
\begin{equation}\label{eq:0011}
    \frac{1}{\sqrt{1+r^2}} (\ket{00} + r \ket{11}),
\end{equation}
where $\ket{0}$ means vacuum and $\ket{1}$ means a single-photon state. First, we analyze required detection efficiencies of this protocol with $r = 1$. We calculate key rates in the same way as our protocols and the direct transmission protocol. For the states in (\ref{eq:0110}) and in (\ref{eq:0011}), we require detection efficiencies of 0.915 and 0.913, respectively. Next, we analyze required detection efficiencies of this protocol optimizing the parameter $r$. Then, we find that the required detection efficiencies for the state in (\ref{eq:0110}) and in (\ref{eq:0011}) are 0.915 and 0.802, respectively. That is, optimizing the parameter $r$ gives significant improvement in terms of required detection efficiencies when we consider the state in (\ref{eq:0011}), while it gives subtle improvement for the state in (\ref{eq:0110}). This phenomenon can be understood as follows~\cite{Alwahaibi2025}. Since the detection efficiency does not affect the vacuum, both of the terms of the state in (\ref{eq:0110}) are affected by the loss, while only the second term is affected for the state in (\ref{eq:0011}). Then, we can reduce the effect of the detection efficiency for the state in (\ref{eq:0011}) by using the parameter $r <1$. In Protocol B, the quantum state distributed among Alice and Bob is similar to this state. However, as shown in Fig.~\ref{fig:DetectionEff}, the required detection efficiency of Protocol B is 0.858, which is larger than 0.802. This can be attributed to the multiphoton components of the two-mode squeezed states. Thus, by eliminating the multiphoton components, we can further relax the required detection efficiency for our long-distance protocol. 
Note that this is possible if high efficiency photon-number-resolving detectors were available.

\bibliography{LDDIQKD}

\end{document}